# Developing Microwave Photonic Temperature Sensors

Arec Jamgochian[1], John Quintavalle[2], Alejandra Torres-Diaz[1], James Filla[2], Gregory F. Strouse[1] and Zeeshan Ahmed[1]

[1]*Thermodynamic Metrology Group, Sensor Science Division, Physical Measurement Laboratory, National Institute of Standards and Technology, Gaithersburg, MD 20899*

[2]*Innovations and Solutions Division, National Institute of Standards and Technology, Gaithersburg, MD 20899*

**Abstract:**

In recent years there has been considerable interest in exploiting the temperature dependence of sapphire whispering gallery mode frequency to develop a mechanically stable, high accuracy temperature sensor. Disk-resonator-based devices have been demonstrated to measure temperature with accuracies of 10 mK or better in the temperature range of 273 K to 373 K. Here we have utilized automated data acquisition and processing to rapidly evaluate a mechanically-stabilized sapphire whispering gallery mode resonator based on a hollow cylinder configuration. Our results indicate the metal support structure contributes to increased loses of the resonator which results in significant reduction in resonator mode's quality factors and lowers the temperature dependent sensor response by $\approx$ 40 %.

**Introduction:**

Since the seminal work of Callender[1] detailing the workings of a platinum resistance thermometer, modern industrial temperature sensing technology has relied on resistance measurement of a thin metal film or wire whose resistance varies with temperature[1, 2]. Though resistance thermometers such as the platinum resistance thermometers (PRT) can routinely measure temperature with uncertainties of 10 mK, sufficient accuracy for most industrial application, they are sensitive to mechanical shock and moisture which causes the sensor resistance to drift over time requiring frequent off-line, expensive and time-consuming calibrations.[2, 3] It is therefore of little surprise that there has been considerable interest in the development of alternatives to resistance thermometry such as photonic temperature sensors [4-6].

In recent years we have focused on the development of sapphire whispering gallery mode resonators (WGMR) for accurate temperature measurement.[4, 7] Although WGMRs have been fabricated with a wide variety of materials including fused quartz and glass spheres, monocrystalline sapphire has emerged as a material of choice as it supports modes with Q-factors of a million.[8] Due to the thermo-optic effect and material thermal expansion, the resonance frequency of sapphire's whispering gallery modes varies with temperature.[4] Considerable effort has been expended in developing a thermally insensitive WGMR for the express purpose of providing a stable frequency source. Realizing that the large temperature dependence of WGM's frequency could be exploited to develop a highly sensitive and accurate temperature sensor, we developed a sapphire WGMR thermometer and demonstrated temperature measurement uncertainty of 10 mK between 273.15 ˚C and 373 ˚C. The WGMR prototype device developed was based on a simple disk shaped resonator.[4] In recent years, spherical and cylindrical resonator geometries have been tested and demonstrated to provide similar measurement capabilities.

Previous experiments have indicated that the WGMR spectrum is sensitive to axial and rotational displacement of the sapphire resonator, highlighting the need for mechanical stabilization of the resonator to enable reproducible measurements. In this study, we evaluate sapphire hollow cylindrical resonators that accommodate a simple two-piece metallic support structure inside the sapphire resonator. Our results indicate the presence of the supporting mechanism suppresses mode quality factors (Q-factors) by a factor of ten and device's temperature sensitivity by ≈ 40 %.

**Experimental:**

*Sapphire Resonators:* Three hollow cylindrical resonators were fabricated for this study. Resonator 1 is 30.0 mm long with outer diameter (OD) of 6.0 mm and inner diameter (ID) of 3.0 mm and the crystal *c*-axis was aligned to the laboratory *z*-axis with ±0.1˚ and its surfaces polished to a mirror finish. Resonator 2 has the same specifications as Resonator 1 except on one end (side B) a 5.0 mm deep counter-bore of OD 4.0 mm was made. Resonator 2 was tested with side B facing the antennas. Resonator 3 differs from resonator 1 in that its surface has two 1 mm deep, 5 mm wide, periodic grooves (50 % duty cycle) etched on; the third groove is only 3 mm wide.

*Cavity:* The resonator cavity fabricated from high conductivity copper consists of a central cylindrical volume (OD 30.0 mm, ID 16.0 mm and length 40 mm) and two circular lids (30.0 mm diameter). All internal surfaces were polished to a mirror finish and gold coated to prevent surface oxidation.

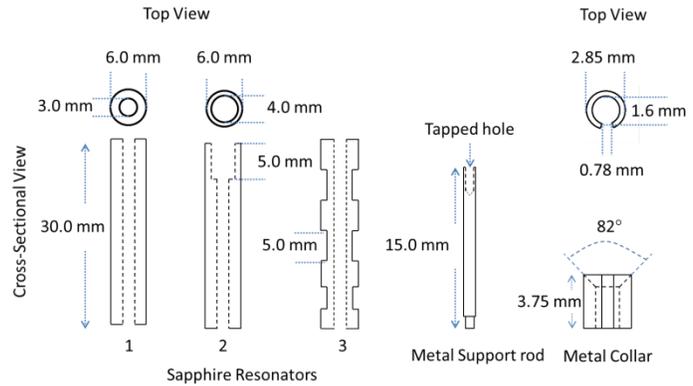

Figure 1: Three hallow cylindrical sapphire resonator and the metal support structure used in this study are show above.

The tube resonator geometry was chosen to allow for the development of a support structure that would enable reproducible re-assembly of the WGMR thermometer. Design consideration is constrained by the need to minimize interference from any conducting parts. Previous experiments suggest that embedding metal parts inside the sapphire resonator can minimize interference, limiting spectral artifacts.[4, 9] As such we designed a rod-collar mechanism for localizing the resonator; this mechanism is entirely contained inside the sapphire tube. The support rod, screwed into the bottom plate is a 15 mm long stainless steel metal rod with a 6.25 mm deep grooved recess for accepting a metal screw. The metal collar, 2.85 mm in diameter and 3.75 mm long, is mated to the support rod using a small screw. Tightening of the screw causes the collar to expand against the inner surface of the sapphire tube. The friction exerted by the collar against the sapphire tube is sufficient to localize it and prevent the resonator from rotating or translating randomly inside the cavity. A 5 mm tall Teflon spacer is placed between the sapphire resonator and the bottom lid to optimize spectral intensity. At 25 °C, upon resonator re-assembly, the center frequency of the highest frequency mode changes by ≈ 5 % ($\Delta T \approx 35$ °C) while the Q-factors change by 75 %. The temperature response ($\delta v/\delta T$) however, is well reproduced ($\Delta T \approx 0.2$ °C). The observed offset in mode frequency likely derives from differing levels of strain imposed by the metallic collar onto the sapphire resonator during re-assembly and needs to standardized in future design iterations.

*Measurement Apparatus:* The sapphire WGMR cavity is thermally cycled in an Isotech[1] drywell. The sapphire WGMR modes are probed using an Agilent N5230A network analyzer[1].

---

[1] Disclaimer: Certain equipment or materials are identified in this paper in order to specify the experimental procedure adequately. Such identification is not intended to imply endorsement by the National Institute of

The experiment is computer controlled using an automated LabView[1] program that is used to cycle the temperature between 25 °C and 75 °C in 5 °C increments. At each temperature the program allows the drywell to reach equilibrium (temperature remains stable to within 0.01 °C for minimum of two minutes at the set point, followed by a 90 min wait time). After the equilibration period, the network analyzer is prompted to record the microwave transmission spectra (S12) of the cavity over the range of 10 MHz to 20 GHz with 25 kHz resolution. During the course of measurement, the temperature of the drywell is continuously monitored using a calibrated PRT. Our results indicate the temperature is maintained to better than ±0.01 °C during spectral acquisition. Following acquisition of spectra, the drywell is prompted to move to the next temperature set point; this pattern is repeated until temperature cycling is completed.

*Data Processing:* The observed spectra are fitted to a spline function which is utilized to identify cavity modes and extract their peak center, width and amplitude at each temperature. Since increase in temperature results in decreased permittivity, the WGMR mode frequency will show a negative temperature dependence, as given by equation 1.

$$\frac{1}{v_o}=\frac{1}{v_o}\{\frac{v_o}{\varepsilon_\perp}\frac{\partial \varepsilon_\perp}{\partial T}+\frac{\partial v_o}{\partial \varepsilon_\parallel}\frac{\partial \varepsilon_\parallel}{\partial T}+\frac{\partial v_o}{\partial L}\frac{\partial L}{\partial T}+\frac{\partial v_o}{\partial a}\frac{\partial a}{\partial T}\} \quad (1)$$

where $v_o$ is the mode frequency, $a$ is the resonator diameter, $L$ is the axial direction, $T$ is temperature, $\varepsilon_\parallel$ is the permittivity in the axial direction, while $\varepsilon_\perp$ is permittivity in radial direaction. We filter out all modes that show a temperature dependence greater than (-350 kHz/°C). The $\delta v/\delta T$ value observed for the empty cavity is only -313 kHz/°C. Thus, choosing a value below this threshold ensures spectral artifacts or cavity modes whose frequency dependence derives chiefly from thermal expansion do not interfere with our spectral interpretation. The resulting list of modes is then scrutinized for identification of WGMR modes and their suitability for applications in thermometry.

**Results and Discussion:** For resonator 1, with metal supporting structure we observe three cavity resonances located at 17.317 GHz, 15.68 GHz and 12.405 GHz. Over the temperature range of 25 °C to 75°C the shift in resonant mode frequency ($\delta v/\delta T$) shows a linear dependence on temperature (Fig 2a) that varies from -358 kHz/°C for 12.4 GHz mode to -704.6 kHz/°C for the 17.3 GHz mode. The observed mode frequency temperature dependence is ≈ 40 % smaller than values observed in previous experiments. Furthermore, the *Q*-factors are at least a factor of 10 smaller than the previously reported cylindrical resonators [2, 7, 9]. The observed decrease in sensor performance likely arises due to coupling of electromagnetic energy between the sapphire resonator and the metal support structure.

The role of metal support structure in degrading the mode *Q*'s is confirmed by replacing the metal support structure with Teflon pieces (collar and support rod). For resonator 1 with Teflon support mechanism we observe five cavity modes located between 15 GHz and 20 GHz. As shown in Fig 2b the mode temperature dependence ($\delta v/\delta T$) varies from -509 kHz/°C to -1217 kHz/°C while the *Q*-values vary from 350 to 9322. The fact that we observe $\delta v/\delta T$ on-par with previously reported disk and rod resonators that were up to three times larger in size [4] bodes well for our future efforts in further reducing the photonic thermometer's size.

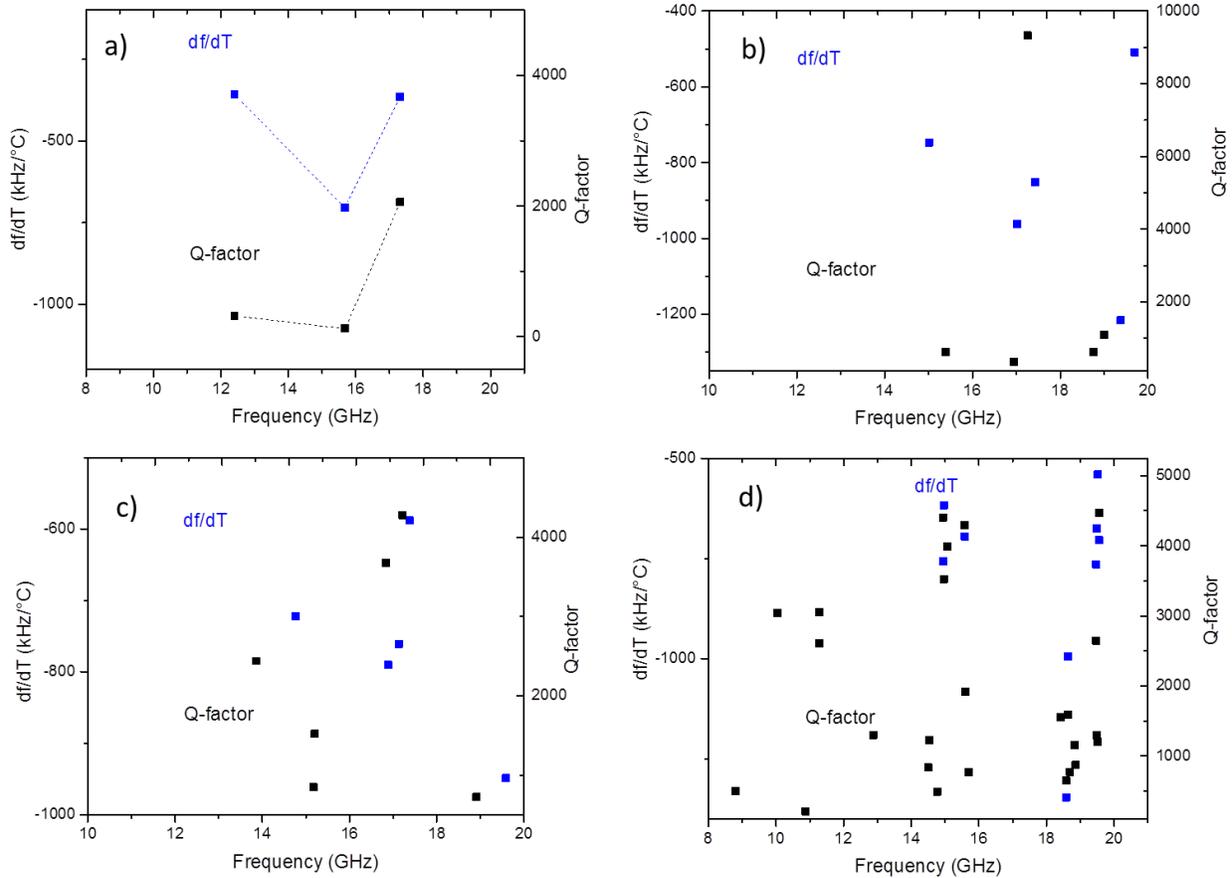

Figure 2: Q-factors and temperature induced frequency shift of resonator 1 modes in the presence of a) metal support structure are depressed relative to when Teflon support structure (b) is used. c) Q-factors and temperature induced frequency shift of resonator 2 modes with Teflon support structure are shown d) Q-factors and temperature induced frequency shift of resonator 3 show a general overall increase in mode density and their Q-factors.

We utilized resonator 2 to test the impact of wall thickness on spectral profile. As shown in Fig 3d, the resonant modes in resonator 2 show a decrease of up to 25 % in their temperature dependent response. This result suggests that shrinking WGMR thermometers to a size comparable to SPRTs (outer diameter 7.5 mm) while providing similar temperature sensitivity, though possible, will require the developemnt of innovative non-metallic support structures. The expected decrease in mode *Q*-factors with shrinking resonator size could potentially be compensated by the use of Bragg resonators.[10] Our preliminary efforts with Bragg structures

(resonator 3) show introduction of Bragg features results in increased mode density and a small increase in overall *Q*-factors of all resonant modes. The temperature dependence is not significantly impacted by the introduction of Bragg features (Fig 2d).

**Summary:** We have utilized an automated data acquisition and processing platform to demonstrate the feasibility of using a simple two-piece metal support structure to stabilize sapphire resonator in a microwave cavity. Our results indicate the metal support structure increases the resonator skin losses, resulting in significantly lower *Q*-factors and temperature dependence. While there is a significant *Q*-factor penalty for using metallic support structures and reducing the size of the sapphire resonator, our results here demonstrate the feasibility of this strategy and suggest the sapphire resonator can be further reduced without significantly compromising the WGMR temperature response. Critical to the development of WGMR thermometry will be the development innovative support structures that can enable reproducible re-assembly of the resonator without significant perturbation of the resonant modes.

**Acknowledgments**: Arec Jamgochian was supported by SURF (Summer Undergraduate Research Fellowship) and Alejandra Diaz was supported by SHIP (Summer High school Intership program).